\documentclass[sigconf]{acmart}
\usepackage{verbatim}
\usepackage{courier}
\usepackage{amsmath}
\usepackage{listings}
\definecolor{mygreen}{rgb}{0,0.6,0}
\usepackage{subcaption}
\usepackage[font=small,labelfont=bf]{caption}
\usepackage{setspace}

\setcopyright{none}
\renewcommand\footnotetextcopyrightpermission[1]{} 
\begin{document}






%

\title{Application-Level Resilience Modeling for HPC Fault Tolerance}
%
\author{Luanzheng Guo}
\affiliation{
\institution{UC Merced}
}
\email{lguo4@ucmerced.edu}

\author{Hanlin He}
\affiliation{
\institution{UC Merced}
}
\email{hhe3@ucmerced.edu}

\author{Dong Li}
\affiliation{
\institution{UC Merced}
}
\email{dli35@ucmerced.edu}

\begin{abstract}
Understanding the application resilience in the presence of faults is critical to address the HPC resilience challenge. Currently we largely rely on random fault injection (RFI) to quantify the application resilience.  However, RFI provides little information on how fault tolerance happens, and RFI results are often not deterministic due to its random nature.
In this paper, we introduce a new methodology to quantify the application resilience. Our methodology is based on the observation that at the application level, the application resilience to faults is due to the application-level fault masking. The application-level fault masking happens because of application-inherent semantics and program constructs. Based on this observation, we analyze application execution information and use a data-oriented approach to model the application resilience. 
We use our model to study how and why HPC applications can (or cannot) tolerate faults. We demonstrate tangible benefits of using the model to direct fault tolerance mechanisms.

\vspace{-10pt}
\end{abstract}

\maketitle

\section{Introduction}
\label{sec:intro}

The high performance computing (HPC) systems are jeopardized by potentially increasing faults
in hardware and software~\cite{asplos15:vilas, fengshui_sc13}. 
Ensuring scientific computing integrity through the correctness of application outcomes in the presence of faults 
remains one of the grand challenges (also known as the resilience challenge) for HPC. 

To address the resilience challenge, we must sufficiently understand the resilience of hardware and applications.
Understanding the resilience of hardware is required to validate the hardware design against a desired failure rate target, 
and improve the efficiency of the protection used to achieve the failure rate~\cite{micro15:steven, micro14:wilkening}.
Understanding the resilience of applications is necessary to determine whether the application execution can
remain correct with fault occurrences and whether we should enforce software-based fault tolerance mechanisms 
(e.g., algorithm-based fault tolerance~\cite{ftcg_ppopp13, tc84_abft} and 
software-based checkpoint/restart~\cite{4-kharbascombining})
to ensure application result fidelity and minimize re-computation cost.
Understanding the resilience of applications is also the key to coordinate software- and hardware-based fault tolerance mechanisms~\cite{CrossLayer_icpp12, abft_ecc:SC13} to improve their efficiency. 

Although researchers in the field have made significant progress on understanding the resilience of hardware based on various methodologies 
(such as (micro)architecture-level simulations~\cite{asplos10:feng, hpca09:li}, beam test on silicon~\cite{aspdac14:cher, sc14:cher}, 
and AVF analysis~\cite{isca05:mukherjee, micro03:mukherjee}),
understanding the resilience of applications largely relies on random fault injection (RFI) at the application level~\cite{europar14:calhoun, mg_ics12, bifit:sc12, sc16_gpu_fault_injection, prdc13:sharma}.  
The random fault injection injects artificial faults into application variables or computation logic,
and obtains statistical comparisons 
of application susceptibility to faults.
To ensure statistical significance and sufficient fault coverage,
this methodology has to perform a large amount of random fault injection
tests. 

However, RFI has a fundamental limitation. 
First, because of the random nature of RFI, it is often difficult to
bound the accuracy of RFI. 
It is difficult to know how many fault injection tests should be performed.
Different numbers of fault injection tests can result in different conclusions on the application resilience (see Section~\ref{sec:bac}).
Although previous work~\cite{date09:leveugle} has shown the possibility of determining the number of fault injection within specific confidence level and error level, it has to estimate the application resilience (e.g., percentage of faults resulting in a crash) before fault injection, which is a priori unknown. 
Also, the fault injection results within the expected error level can still be 
different from the real application resilience~\cite{date09:leveugle}.
Second, RFI gives us little knowledge on how and where faults are tolerated. 
Having such knowledge is important to determine where to enforce fault tolerance mechanisms.

The limitation of RFI creates a major obstacle to implement efficient fault tolerance mechanisms.
Many fault tolerance strategies, such as selective protection~\cite{sc14:elliott, snl_tr11:Hoemmen} 
and cross-layer protection~\cite{CrossLayer_icpp12, abft_ecc:SC13}, will be difficult
to be enforced without sufficient information on the application resilience.
Hence, we desire a 
methodology alternative to RFI to understand the application resilience.

In this paper, we introduce a fundamentally new methodology to quantify and model the application resilience. Our methodology is based on an observation that at the application level, the application resilience to faults is due to application-level fault masking.
The application-level fault masking happens because of application-inherent semantics and program constructs. For example, a corrupted bit in a data structure could be overwritten by an assignment operation, hence does not cause
incorrect application outcomes; a corrupted bit of a molecular representation in the Monte Carlo method-based simulation to
study molecular dynamics may not matter to application outcomes, because of the statistical nature of the simulation.

Based on the above observation, the quantification of the application resilience at the application level
is equivalent to quantifying fault masking at the application level.
By analyzing the application execution information (e.g., the architecture-independent, LLVM~\cite{llvm_lrm} dynamic instruction trace), we can accurately capture those application-level fault masking events, and provide insightful analysis on whether there is any fault tolerance and how it happens.

In essence, RFI attempts to opportunistically capture those fault masking events:
an RFI test without causing incorrect application outcomes 
captures one or more fault masking events, and is counted 
to calculate the success rate (or failure rate) of all fault injection tests.
However, the random nature of RFI can miss or redundantly count fault masking events.
Depending on when and where the RFI tests happen, different RFI tests 
can result in different results when evaluating the application resilience.
Our methodology avoids the randomness, hence avoids the limitation of
the traditional RFI.   

To capture the application-level fault masking, we classify common fault masking events
into three classes: operation-level fault masking, fault masking due to fault propagation,
and algorithm-level fault masking. The operation-level fault masking 
takes effect at individual operations (e.g., arithmetic computation and assignment), and broadly includes
value overwriting, logical and comparison operations, and value shadowing.
The fault masking due to fault propagation takes effect across operations,
and requires tracking data flows between operations of the application. 
The algorithm-level fault masking manifests at the end of the application execution.
To identify the algorithm-level fault masking, 
we introduce deterministic fault injection, and
treat the application as a black box without requiring 
detailed knowledge of the algorithm/application internal mechanisms and semantics. 

Our application-level resilience modeling opens new opportunities to
examine applications and evaluate the effectiveness of application-level fault 
tolerance mechanisms. It is applicable to a number of use cases to address the resilience challenge, such as code optimization and algorithm choice.
In summary, this paper makes the following contributions:

\begin{itemize}
\item
We introduce a novel methodology to model the application resilience.
Our methodology avoids the randomness inherent in the traditional random fault injection,
and brings deterministic and insightful quantification of the application resilience, which is unprecedented.


\item
We comprehensively investigate application-level fault masking events and classify them.
Our investigation answers a primitive question: why can an application tolerate faults at the application level? 
Answering this question is fundamental for enabling resilient applications for HPC and designing efficient fault tolerance mechanisms.

\item
We introduce a set of techniques to identify fault masking events. 
Furthermore, we develop a modeling tool based on our modeling methodology and techniques.
The tool is highly configurable and extensible, making the modeling work 
practical and flexible. We apply our tool to representative, computational algorithms and two scientific applications.
We reveal how fault masking typically happens in HPC applications. 

\item 
Using one case study to determine the deployment of fault tolerance mechanisms, we demonstrate tangible benefits of using a model-driven approach to direct fault tolerance designs for HPC applications.

\end{itemize}
\section{Background}
\label{sec:bac}
In this section, we introduce the basic fault model for the application-level resilience modeling.
We also give an introductory description for fault masking at the application level. 
 
\subsection{Fault Model}

In this paper, we model the impact of any fault on applications at the application level, 
and do not consider where the faults are originally generated (register, main memory, cache, etc). 
As long as the faults manifest and are propagated to the application level,
we will model if the application is resilient to those faults.  
In addition, we consider single-bit and spatial multi-bit transient faults, because those fault patterns are the most common
ones and impose a more significant threat than others~\cite{asplos15:vilas, fengshui_sc13, micro14:wilkening}. 

In terms of the impact of faults on applications, we focus on the execution correctness. 
We define the execution correctness at the application level (in contrast to the architecture level in the related work~\cite{isca05:mukherjee, micro03:mukherjee}).
An application's execution is deemed correct as long as the outcome it produces is \textit{acceptable}.
Depending on the notion of the outcome acceptance, the correctness can refer to precise numerical integrity 
(e.g., the numerical outcome of a multiplication operation must be precise), 
or refer to satisfying a minimum fidelity threshold (e.g., the outcome of an iterative solver must meet certain convergence thresholds).    

\subsection{Randomness of Traditional Fault Injection}
The traditional fault injection injects faults randomly into the application.
The randomness results in uncertainty in the fault injection result. 
To motivate our modeling work, we study the randomness of traditional random fault injection.
We leverage an LLVM-based fault injection tool~\cite{europar14:calhoun} and study several  benchmarks from NPB benchmark suite 3.3.1~\cite{nas}. For each fault injection test,
this tool randomly selects an instruction and then randomly flips a bit in the output operand of the instruction. We use single-bit flip for fault injection tests,
similar to the existing work~\cite{europar14:calhoun, mg_ics12, bifit:sc12,  prdc13:sharma}.
We use the statistical approach in~\cite{date09:leveugle} to
quantify error level with 95\% confidence level.
A priori estimation of the application resilience is 0.5 suggested by~\cite{date09:leveugle}.
We do ten sets of fault injection tests.
The number of fault injection tests in the ten sets ranges
from 1000 to 10,000 with a stride of 1000.

\begin{figure}
	\begin{center}
		\includegraphics[height=0.2\textheight]{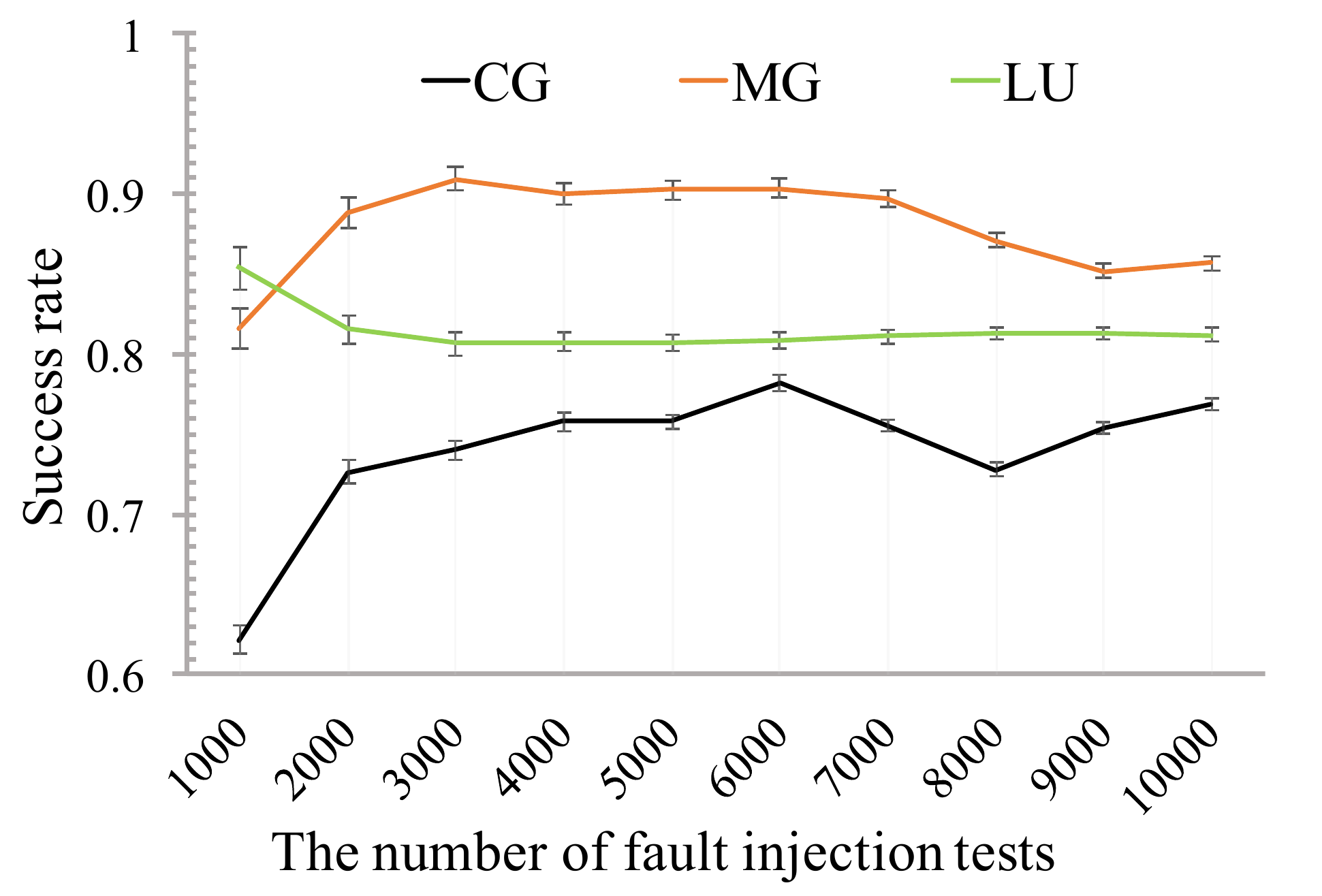} 
		\vspace{-10pt}
		\caption{The random fault injection results with margins of error for CG, MG, and LU (CLASS B) with 95\% confidence level.}
		\label{fig:fi_randomness}
		\vspace{-15pt}
	\end{center}
\end{figure}

Figure~\ref{fig:fi_randomness} shows the results, and uses the success rate of fault injection to evaluate the application resilience, similar to~\cite{ics08:bronevetsky, 
bifit:sc12,  prdc13:sharma}.
The success rate of fault injection is defined as follows: Among $N$ fault injection tests, if $M$ of them have correct application outcomes, then the
success rate is calculated as $M/N$. 
A higher value of the success rate indicates that the application is more resilient to faults.
Within the figure, we also show the margin of error (i.e., error level) for the success rate based on~\cite{date09:leveugle}.

The figure reveals that the fault injection result is sensitive to the number of fault injection tests even with the consideration of error level and confidence level.
For CG, the fault injection results are 62\% for 1000 fault
inject tests and 78\% for 6000 fault injection tests.
Furthermore, when the number of fault injection tests is 1000, we find that
LU is slightly more resilient than MG (0.81 for MG vs.
0.86 for LU). However, when the number of fault injection
tests is 3000, MG is more resilient than LU (0.91 for MG vs. 0.81 for LU). 
We make totally different conclusions when comparing MG and LU.
This observation is a clear demonstration of the randomness of
using the traditional fault injection to study the application resilience.
Such randomness comes from the limitation of the statistical approach,
in particular, a priori estimation of the application resilience, 
limited confidence level, and inability to capture some fault masking events. 
We must have a new methodology.

\subsection{Fault Masking}
Fault masking can happen at the application level and hardware level.
The application level fault masking happens because of application inherent semantics and program constructs.
The hardware level fault masking happens because a fault does not corrupt the precise semantics of hardware.
For example, branch mis-speculated states due to branch prediction or speculative memory disambiguation, if corrupted, will not cause incorrect outcome.

Our resilience modeling focuses on the application-level fault masking, and does not include fault masking at the hardware level. 
Furthermore, we use a data-oriented approach, and focus on \textit{fault masking happened in individual data objects}. 
In other words, we consider that when a fault (data corruption) happens in a data object, whether the fault can be masked.
A data object can be, for example, a matrix in matrix multiplication or a tree structure in the Barnes-Hut N-body simulation.
Using the data-oriented approach is beneficial for the resilience research, because HPC applications
are often characterized by a large amount of data objects, and the application outcomes are typically stored in data objects.
Furthermore, many popular fault tolerance mechanisms are designed to protect data objects (e.g., application-level checkpoint/restart mechanisms
and many algorithm-based fault tolerance methods~\cite{ftcg_ppopp13, ft_lu_hpdc13, jcs13:wu}). Quantifying the resilience of data objects
can greatly benefit the designs of these fault tolerance mechanisms.
For example, if a data object is resilient, then we do not need to apply these mechanisms to protect it, which will improve performance and energy efficiency. In Section~\ref{sec:case_study}, we have a case study to further demonstrate the benefits of our approach.



\section{Application-Level Resilience \\ Modeling}
\label{sec:modeling}
This section describes our modeling methodology. 
We start with a classification of the application-level fault masking,
and then introduce a metric and investigate how to use it
to quantify the application resilience. 

\subsection{General Description}
\label{sec:general_bg}
Application-level fault masking has various representations.  
Listing~\ref{fig:general_desc} gives an example to illustrate the application-level fault masking. 
In this example, we focus on a data object, $par\_A$, which is a sparse matrix with 1$K$ non-zero data elements. We study \textit{fault masking happened in this data object}. $par\_A$ is involved in 4 statements (Lines 6, 7, 9 and 13). 

In this example, the statement at Line 6 has a fault masking event:
if a fault happened at a data element $par\_A[0].data$ of the target data object ($par\_A$), the fault can be overwritten by an assignment operation.
The statement at Line 7 has no explicit fault masking event happened in the target data
object, but if a fault at a data element $par\_A[2].data$ occurs, the fault is propagated to $c$ by multiplication and assignment operations.
At Line 9, assuming that the value of $c$ is much smaller than the value of a variable $GIANT$, 
the impact of the corrupted $c$ on the application outcome is ignorable.
Hence, the fault propagated from Line 7 to Line 9 can be indirectly masked.

\begin{lstlisting}[label={fig:general_desc}, caption={An example code to show application-level fault masking}]
void func (Matrix *par_A, Vector *par_b, Vector *par_x) {
	// the data object par_A has 1K data elements;
    float c=0.0;
    
    // pre-processing par_A
    par_A[0].data=sqrt(initInfo);
    c=par_A[2].data*2;
    if (c>THR) {
    	par_A[4].data=c+GIANT; // GIANT >> c
    }
    
    // using the algebraic multi-grid solve
    AMG_Sover(par_A, par_b, par_x);
}
\end{lstlisting}

At Line 9, there is also an explicit fault masking event (i.e., fault overwritten by an assignment operation) for $par\_A[4].data$ if a fault happens in $par\_A[4].data$. This fault masking is similar to the one at Line 6.
At Line 13, there is an invocation of an algebraic multi-grid solver (AMG)
that can tolerate faults in the matrix because of the algorithm-level semantics of AMG (particularly, AMG's iterative, multilevel structure~\cite{mg_ics12}).

This example reveals many interesting facts.
In essence, a program can be regarded as a combination of data objects and
operations performed on the data objects.
An operation refers to the arithmetic computation, assignment, logical and comparison operations,  
or an invocation of an algorithm implementation (e.g.,  a multigrid solver, a conjugate gradient method, or a Monte Carlo simulation).  
An operation may inherently come with fault masking effects, exemplified at Line 6 (fault overwritten);
An operation may propagate faults, exemplified at Line 7. 
Different operations have different fault masking effects, and hence
impact the application outcome differently.
Different applications can have different operations because of
algorithm implementation and compiler optimization, hence the
applications can have different application-level resilience.

Based on the above discussion, we classify application-level fault masking 
into three classes.

(1) \textbf{Operation-level fault masking.} At individual operations, a fault happened in a data object is masked because of the semantics of the operations. Line 6 in
Listing~\ref{fig:general_desc} is an example.

(2) \textbf{Fault masking due to fault propagation.} 
Some fault masking events are implicit and have to be identified beyond a single operation. 
In particular, a corrupted bit in a data object is not masked in the current operation (e.g., Line 7 in 
Listing~\ref{fig:general_desc}),
but the fault is propagated to another data object and masked in another operation (e.g., Line 9).
Note that simply relying on the operation-level analysis without the fault propagation analysis is not sufficient to recognize these fault masking events.

(3) \textbf{Algorithm-level fault masking.}
Identification of some fault masking events happened in a data object must include algorithm-level information.
The identification of those events is beyond the first two classes.
Examples of such events include 
the multigrid solver~\cite{mg_ics12}, some iterative methods~\cite{2-shantharam2011characterizing}, and certain sorting algorithm~\cite{prdc13:sharma}.  
Furthermore, some application domains, such as image processing and machine learning~\cite{isca07:li}, can also tolerate faults because of less 
strict requirements on the correctness of data values. 

In general, the first two classes are caused by program constructs, and the third class is caused by algorithm semantics. Due to the random nature, 
the traditional random fault injection may omit some fault masking events, or capture them multiple times.
Relying on analytical modeling, we can avoid or control the randomness of the fault injection, hence greatly improve resilience evaluation. 

Our resilience modeling is analytical, and 
relies on the quantification of the above application-level fault masking events happened on data objects.
We create a new metric to quantify the application-level resilience at \textit{data objects}, and introduce methods to measure the metric based on the above classification of fault masking events.

\vspace{-10pt}
\subsection{aDVF: An Application-Level Resilience \\ Metric}
\label{sec:metric}
To quantify the resilience of a data object due to fault masking events, we could simply count the number of fault masking events
that happen to the target data object. 
However, a direct resilience comparison between data objects in terms of the number of fault masking events cannot provide meaningful quantification of the resilience of data objects. 
For example, a data object 
may be involved in a lot of fault masking events, 
but this does not necessarily mean this data object is more resilient to faults
than other data objects with fewer fault masking events, because the fault masking events of this data object 
can come from a few repeated operations, and the number of fault masking events is accumulated throughout application execution;
This data object could be not resilient, if most of other operations for this data object do not have fault masking. 
Hence, the key to quantifying the resilience of a data object is
to quantify \textit{how often} fault masking happens to the data object.
We introduce a new metric, \textit{aDVF} (i.e., the application-level Data Vulnerability Factor), 
to quantify application-inherent resilience due to fault masking in data objects. aDVF is defined as follows. 

For an operation performed on a data element of a data object, we reason that if a fault happens at the data element in this operation, 
the application outcome could or could not remain correct in terms of the outcome value and application semantics. 
If the fault does not cause an incorrect application outcome,
then a fault masking event happens to the target data object.
A single operation can operate on one or more data elements of the target data object. 
For a specific operation, aDVF of the target data object is defined as the total number of fault masking events divided by the total number of data elements of the target data object operated on by the operation.

For example, an assignment operation $a[1] = w$ 
happens to a data object, the array $a$.
This operation involves one data element ($a[1]$) of the data object $a$.
We calculate aDVF for the target data object $a$ in this operation as follows.
If a fault happens to $a[1]$, we deduce that 
the erroneous $a[1]$ does not impact application correctness and the fault in $a[1]$ is always masked. Hence, the number of fault masking events for
the target data object $a$ in this operation is 1. Also, the total number of data elements involved in the operation is 1.
Hence, the aDVF value for the target data object in this addition operation is $1/1=1$.

Based on the above discussion, the definition of aDVF for a data object $X$ in an operation ($aDVF^{X}_{op}$)
is formulated in Equation~\ref{eq:dvf}, where 
$x_i$ is a data element of the target data object $X$, and $m$ is the number of data elements operated on by the operation;
$f$ is a function to count fault masking events happened on a data element. 
\vspace{-1pt}
\begin{equation} 
\label{eq:dvf}
\footnotesize
	aDVF^{X}_{op} = \sum_{i=0}^{m-1}f(x_i)/m
\end{equation}
\vspace{-5pt}

The calculation of aDVF for a code segment is similar to the above for an operation, except that
$m$ is the total number of $x$
involved in all 
operations of the code segment. 
To further explain it, we use as an example
a code segment from LU benchmark in SNU\_NPB benchmark suite 1.0.3 (a C-based implementation of the Fortran-based NPB) shown in 
Listing~\ref{fig:advf_example}.

\begin{lstlisting}[label={fig:advf_example}, caption={A code segment from LU.}]
void l2norm(int ldx, int ldy, int ldz, int nx0, \
	int ny0, int nz0, int ist, int iend, int jst, \
    int jend, double v[][ldy/2*2+1][ldx/2*2+1][5], \
    double sum[5])
{
	int i, j, k, m;
    for (m=0;m<5;m++) //the first loop
    	sum[m]=0.0;  //Statement A
    
    for (k=1;k<nz0-1;k++){  //the second loop
    	for (j=jst;j<jend;j++){
        	for (i=ist;i<iend;i++){
            	for (m=0;m<5,m++){
            		sum[m]=sum[m]+v[k][j][i][m]  \
                    	*v[k][j][i][m]; //Statement B
                }
            }
        }
    }
    
    for (m=0;m<5;m++){  //the third loop
    	sum[m]=sqrt(sum[m]/((nx0-2)*  \
        	(ny0-2)*(nz0-2))); //Statement C
    }
} 
\end{lstlisting}

\textbf{An example from LU.} We calculate aDVF for the array $sum[]$. 
Statement $A$ has an assignment operation involving one data element ($sum[m]$) and one fault masking event (i.e., if a fault happens to $sum[m]$, the fault is overwritten by the assignment). Considering that there are five iterations in the first loop ($iter_{num1} = 5$), there are 5 fault masking events happened in 5 data elements of $sum[]$

Statement B has two operations related to $sum[]$ (i.e., an assignment and an addition). The assignment operation involves one data element ($sum[m]$) and one fault masking; the addition operation involves one data element ($sum[m]$) and one potential fault masking (i.e., certain corruptions in $sum[m]$ can be ignored, if ($v[k][j][i][m]*v[k][j][i][m]$) is significantly larger than $sum[m]$). This potential fault masking is counted as $r^\prime$ ($0 \leq r^\prime \leq 1$), depending on where a corruption happens in $sum[m]$ and fault propagation analysis result (see Sections~\ref{sec:statement_analysis} and~\ref{sec:impl} for further discussion). 
Considering the loop structure, there are ($(1+r^\prime) * iter_{num2}$) fault masking events happened in ($2 * iter_{num2}$) elements of $sum[]$, where ``1'' and ``$r^\prime$'' come from the assignment and addition operations respectively. $iter_{num2}$ is the number of iterations in the second loop, which is equal to ($(nz0-2)*(jend-jst)*(iend-jst)*5$).

\begin{figure}
	\centering
	\includegraphics[height=0.15\textheight, width=0.48\textwidth]{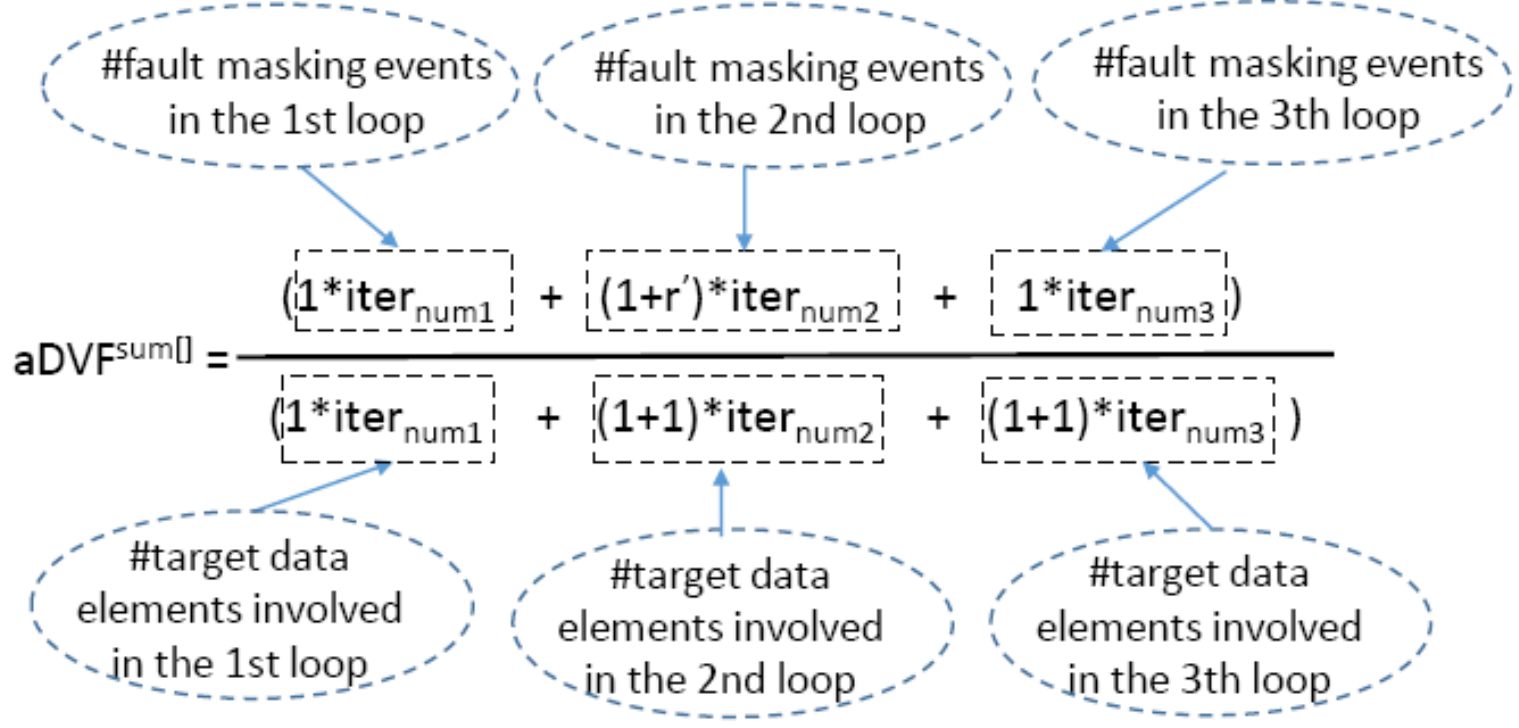}
	\vspace{-8pt}
	\caption{Calculating aDVF for a target data object, the array \textit{sum}[] in a code segment from LU. In the figure, $iter_{num1}=5, iter_{num3}=5$ and $iter_{num2} = (nz0-2)*(jend-jst)*(iend-ist)*5.$}
	\label{fig:advf_cal}
	\vspace{-15pt}
\end{figure}

Statement C has two operations 
related to $sum[]$ (i.e., an assignment and a division), but only the assignment operation has fault masking.
Considering that there are 5 iterations in the third loop ($iter_{num3} = 5$), there are 5 fault masking events happened on 5 data elements of the target data object in the third loop. In 
summary, the aDVF calculation for $sum[]$ is shown in Figure~\ref{fig:advf_cal}.  

To calculate aDVF for a data object, we must rely on effective identification and counting of fault masking events (i.e., the function $f$).
In Sections~\ref{sec:statement_analysis},~\ref{sec:fault_propagation_analysis} and ~\ref{sec:algo_analysis}, 
we introduce a series of counting methods based on the classification of fault masking events. 

\subsection{Operation-Level Analysis}
\label{sec:statement_analysis}
To identify fault masking events at the operation level, we analyze 
all possible operations. 
In particular, we analyze 
architecture-independent, LLVM instructions 
and characterize them based on the instruction result sensitivity to corrupted operands. We classify the operation-level fault masking as follows. 

(1) \textbf{Value overwriting}.  
An operation writes a new value into the target data object, 
and the fault in the target data object is masked. 
For example, the store operation overwrites the fault in the store destination. 
We also include \textit{trunc} and bit-shifting operations into this category, because the fault can be truncated or shifted away in those operations.

(2) \textbf{Logical and comparison operations}.
If a fault in the target data object does not
change the correctness of logical and comparison operations, the fault is masked.  
Examples of such operations 
include logical \textit{AND} and the predicate expression in a \textit{switch} statement.

(3) \textbf{Value shadowing}.
If the corrupted data value in an operand of an operation 
is shadowed by other correct operands involved in the operation,
then the corrupted data has an 
ignorable impact on the correctness of the operation.
The addition operation at Line 9 in Figure~\ref{fig:general_desc} is such an example. 
We can find many other examples, such as arithmetic multiplication. 
The effectiveness of value shadowing is coupled with the application semantics.  
An operation of $1000+0.0012$ can be treated as equal to $1000+0.0011$ without impacting the execution correctness of application,
while such tiny difference in the two data values may be intolerable in a different application. We will discuss how to identify value shadowing in details in Section~\ref{sec:impl}.

Since we focus on the \textit{application}-level resilience modeling,
we do not consider those LLVM instructions that do not have
directly corresponding operations at the application statement level for fault masking analysis. 
Examples of those instructions 
include \textit{getelementptr} (getting the address of a sub-element of an aggregate data structure)
and \textit{phi} (implementing the $\phi$ node in the SSA graph~\cite{llvm_lrm}).

The effectiveness of the operation-level fault masking heavily relies on the fault pattern.
The fault pattern is defined by how fault bits are distributed within
a faulty data element (e.g., single-bit vs. spatial multiple-bit, least significant bit vs. most significant bit, mantissa vs. exponent).
To account for the effects of various fault patterns, an ideal method to count fault masking events 
would be to collect fault patterns in a production environment during a sufficiently long time period, and then use the realistic fault patterns to guide fault masking analysis. 
However, this method is not always practical. 
In the practice of our resilience modeling, we enumerate possible fault patterns for a given operation, 
and derive the existence of fault masking for each fault pattern.
Suppose there are $n$ fault patterns, and $m$ ($0 \leq m \leq n$) of which have fault masking happened.
Then, the number of fault masking events is calculated as $m$/$n$,
which is a statistical quantification of possible fault masking.
Using this statistical quantification means that the number of fault masking events can be non-integer. 
We employ the above enumeration analysis to model fault masking for single-bit faults in our evaluation section, but the method of the enumeration analysis can be applied to analyze all fault patterns.
\vspace{-10pt}

\subsection{Fault Propagation Analysis}
\label{sec:fault_propagation_analysis}
At an operation performed on the target data object, 
if a fault happened in the target data object cannot be masked at the current operation, 
then we use the fault propagation analysis to track whether the corrupted data
is propagated to other data object(s) and the faults (including the original one and the new ones propagated to other data object(s)) 
are masked in the successor operations.
If all of the faults are masked, then we claim that the original fault happened in the target data object is masked.   

For the fault propagation analysis, a big challenge is to 
track all contaminated data which can quickly increase as the fault propagates. 
Tracking a large number of contaminated data objects largely increases
analysis time and memory usage. 
To handle the above fault propagation problem, we avoid tracking fault propagation along a long chain of operations
to accelerate the analysis.
We introduce an optimization technique to avoid long tracking.

\textbf{Optimization: bounding propagation path.} 
We take a sample of the whole fault propagation path.
In particular, we only track the first $k$ operations. 
If the original fault and the new faults propagated to other data object(s)
cannot be masked within the first $k$ operations, then we conclude that 
all of the faults will not be masked after the $k$ operations.

This method, as an analysis approximation, could introduce analysis inaccuracy because of the sampling nature of the method. 
However, for a fault that propagates to a large amount of data objects, 
bounding the fault propagation path does not cause inaccurate analysis, because given a large amount of corrupted data, it is highly unlikely that all faults are masked, and
making a conclusion of no fault masking is correct in most cases.
In the evaluation section, we explore the sensitivity of analysis correctness to the length of the fault propagation path (i.e., $k$). We find that setting the propagation path to 10 is
good to achieve accurate resilience modeling in most cases (87.5\% of all cases). Setting it to 50 is good for all cases.

\vspace{-10pt}
\subsection{Algorithm-Level Analysis}
\label{sec:algo_analysis}
Identifying the algorithm-level fault masking demands domain and algorithm knowledge.  
In our resilience modeling, we want to minimize the usage of domain and algorithm knowledge, such that
the modeling methodology can be general across different domains.


We use the following strategy to identify the algorithm-level fault masking (see the next paragraph). 
Furthermore, the user can optionally provide a threshold to indicate a satisfiable solution quality.
For example, for an iterative solver such as conjugate gradient and successive over relaxation,
this threshold can be the threshold that governs the convergence of the algorithms.
For the support vector machine algorithm (an artificial intelligence algorithm), this threshold can be a percentage (e.g., 5\%)
of result difference after the fault corruption.
Working hand-in-hand, the strategy (see the next paragraph) and user-defined threshold treat the algorithm as a black box without
requiring detailed knowledge of the algorithm internal mechanisms and semantics. 
We explain the strategy as follows.

\textbf{A practical strategy for algorithm-level analysis: deterministic fault injection.}
The traditional random fault injection treats the program as a black-box. 
Hence, using the traditional random fault injection could be an effective tool to identify the algorithm-level fault masking.
However, to avoid the limitation of the traditional random fault injection (i.e., randomness), 
we use the operation-level analysis and fault propagation analysis to guide fault injection, 
without blindly enforcing fault injection as the traditional method.
In particular, when we cannot determine whether a fault masking can happen in the target data object for an operation $op$ because of fault propagation,  we track fault propagation until 
reaching the boundary of the fault propagation analysis.
If we still cannot determine fault masking at the boundary, then  
we inject a fault into the target data object in $op$, 
and then run the application to completion. 
If the application result is different from the fault-free result, 
but does not go beyond the user-defined threshold, we claim that the algorithm-level fault masking takes effect. 

\textbf{Discussion: coupling between fault propagation and algorithm level analysis.}
The fault propagation analysis and algorithm-level analysis are tightly coupled.
If we reach the boundary of the fault propagation analysis and cannot determine fault masking, we use the algorithm-level analysis. 
However, by doing this, 
some of the fault masking events due to the fault propagation and operation-level fault masking after the boundary may be accounted as algorithm-level fault masking.
Although this mis-counting will not impact the correctness of aDVF value, it would overestimate the algorithm-level fault masking.

The fundamental reason for the above overestimation is that we bound the boundary of the fault propagation analysis.
However, our study (Section~\ref{sec:eval_sen}) reveals that we can have very good modeling
accuracy on our count of the algorithm level fault masking,
even if we set the boundary of the fault propagation analysis.
The reason is as follows. 
After the boundary of the fault propagation analysis, the fault is widely propagated, and
the chance to mask all propagated faults by the operation-level fault masking is extremely low.
In fact, in our tests, we found that even if we use a longer fault propagation path to identify fault masking, we are not able to find more fault masking based on the fault propagation analysis.
Hence, as long as the threshold is sufficiently large (e.g., 10), 
we do not overestimate the algorithm-level fault masking. 
\vspace{-10pt}	 
\section{Implementation}
\label{sec:impl}

\begin{figure*}
  \begin{center}
  \includegraphics[height=0.13\textheight,keepaspectratio]{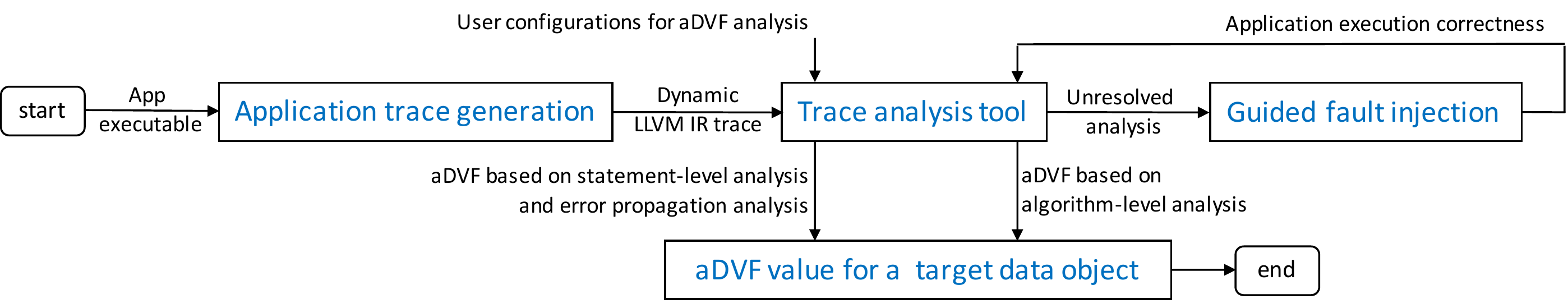}
  \vspace{-5pt}
  \caption{ARAT, a tool for application-level resilience modeling based on the aDVF analysis}
  \label{fig:tool_framework}
  \end{center}
  \vspace{-15pt}
\end{figure*}

To calculate the aDVF value for a data object, 
we develop a tool, named~\textit{ARAT} (standing for \textit{A}pplication-level \textit{R}esilience \textit{A}nalysis \textit{T}ool). 
Figure~\ref{fig:tool_framework} shows the tool framework.
ARAT has three components: an application trace generator, a trace analysis tool, and a deterministic fault injector.

The \textbf{application trace generator} is an LLVM instrumentation pass to generate a dynamic LLVM IR trace.
LLVM IR is architecture independent, and each instruction in the IR trace corresponds to one operation.
The trace includes dynamic register values and memory addresses referenced in each operation.
The current trace generator is based on a third-party tool~\cite{ispass13:shao}, but with some extensions 
for the deterministic fault injection 
and Phi instruction processing to identify ambiguous branches.

The \textbf{trace analysis tool} is the core of ARAT. Using an application trace as input, the tool calculates the aDVF value of a given data object.
In particular, the trace analysis tool conducts the operation-level and fault propagation analysis. 
Also, for those unresolved fault propagation analyses that reach the boundary of the fault propagation analysis,   
the trace analysis tool will output 
a set of fault injection information for the deterministic fault injection. Such information includes dynamic instruction IDs, IDs of the operands that reference the target data object, and the bit locations of the operands that have undetermined fault masking.
After the fault injection results (i.e., the existence of algorithm-level fault masking or not) are available from the deterministic fault injector,
we re-run the trace analysis tool, and use the fault injection results to address the unresolved analyses and update the aDVF calculation. 

For the fault propagation analysis, we associate data semantics (the data object name) with the data values in registers,
such that we can identify the data of the target data object in registers. 
This is necessary to analyze fault propagation.
To associate data semantics with the data in registers, 
ARAT tracks the register allocation when analyzing the trace, such that we can know at any moment which registers have the data of the target data object. 

For the value shadowing analysis to determine which bits can have their bit flips masked, we ask users to provide a set of value shadowing thresholds, each of which defines
a boundary (either upper bound or lower bound) of valid data values for a data element of the target data object. 
Only those bit positions whose bit flips result in a valid data value
are determined to have the fault masking of value shadowing.
If users cannot provide such thresholds, then we will perform deterministic fault injection test for each bit of the data element of the target data object to determine the effect of bit flip on the application outcome.
To accelerate the value shadowing analysis, we further introduce
a series of optimization techniques, such as (1) using the 
deterministic fault injection results for higher-order bits to deduce
the fault injection results for lower-order bits; (2) leveraging
iterative structure of the application;
and (3) analysis parallelization. 
Those implementation details can be found in our technical report~\cite{resilience_modeling:tr}. 

The trace analysis tool is configurable and extensible. 
It gives the user flexibility to control the trace analysis. 
For example, the user can define a maximum fault propagation length for the fault propagation analysis; 
the user can also configure fault patterns for analysis. 
Based on the user configuration, the tool can enumerate all fault patterns during the analysis or just examine one specific pattern.
To make the trace analysis tool extensible for future improvement,
the tool also exposes APIs that allow users to hook up new
techniques to identify fault masking and optimize analysis.

The \textbf{deterministic fault injector} is a tool to capture the algorithm level fault masking
for the trace analysis tool.
The input to the deterministic fault injector is a list of fault injection points
generated by the trace analysis tool for those unresolved fault masking analyses.
Each fault injection point includes a dynamic instruction ID, 
the ID of the operand that references the target data object, and a specific bit-position of the operand for bit flipping (i.e., the fault injection).
The bit-positions of the operand for bit flipping are determined after the value shadowing analysis. 

Similar to the application trace generation, the deterministic fault injector is also based on the LLVM instrumentation. 
We use the LLVM instrumentation to count dynamic instructions and trigger bit flips. 
After the LLVM instrumentation, the application execution will trigger bit flip when a fault injection point is encountered.

\section{Evaluation}
\label{sec:evaluation}
\begin{table*}[!t]
\begin{center}
\caption {Benchmarks and applications for the study of the application-level resilience}
\vspace{-5pt}
\label{tab:benchmark}
\tiny
\begin{tabular}{|p{1.7cm}|p{7.5cm}|p{4cm}|p{2.5cm}|}
\hline
\textbf{Name} 	& \textbf{Benchmark description} 		& \textbf{Execution phase for evaluation}  			& \textbf{Target data objects}             \\ \hline \hline
CG (NPB)             & Conjugate Gradient, irregular memory access (input class S)   & The routine conj\_grad in the main computation loop  & The arrays $r$ and $colidx$     \\\hline
MG (NPB)    	       & Multi-Grid on a sequence of meshes (input class S)             & The routine mg3P in the main computation loop & The arrays $u$ and $r$ 	\\ \hline
FT (NPB)             & Discrete 3D fast Fourier Transform (input class S)            & The routine fftXYZ in the main computation loop  & The arrays $plane$ and $exp1$    \\ \hline
BT (NPB)             & Block Tri-diagonal solver (input class S)         		& The routine x\_solve in the main computation loop & The arrays $grid\_points$ and $u$	\\ \hline
SP (NPB)             & Scalar Penta-diagonal solver (input class S)         		& The routine x\_solve in the main computation loop & The arrays $rhoi$ and $grid\_points$  \\ \hline
LU (NPB)            & Lower-Upper Gauss-Seidel solver (input class S)        	& The routine ssor 	& The arrays $u$ and $rsd$ \\ \hline \hline
LULESH~\cite{IPDPS13:LULESH} & Unstructured Lagrangian explicit shock hydrodynamics (input 5x5x5) & 
The routine CalcMonotonicQRegionForElems 
& The arrays $m\_elemBC$ and $m\_delv\_zeta$ \\ \hline
AMG2013~\cite{anm02:amg} & An algebraic multigrid solver for linear systems arising from problems on unstructured grids (we use  GMRES(10) with AMG preconditioner). We use a compact version from LLNL with input matrix $aniso$. & The routine hypre\_GMRESSolve & The arrays $ipiv$ and $A$   \\ \hline
\end{tabular}
\end{center}
\vspace{-5pt}
\end{table*}

We study 12 data objects from six benchmarks of the NAS parallel benchmark (NPB) suite (we use SNU\_NPB-1.0.3) and 4 data objects from two scientific applications. 
Those data objects are chosen to be representative: they have various data access patterns and participate in various execution phases.  
For those benchmarks and applications, we use their default compiler options, and use gcc 4.7.3 and LLVM 3.4.2 for trace generation.
To count the algorithm-level fault masking, we use the default convergence thresholds (or the fault tolerance levels) for those benchmarks.
Table~\ref{tab:benchmark} gives 
detailed information on the benchmarks and applications.
The maximum fault propagation path for aDVF analysis is set to 10 by default.

\subsection{Resilience Modeling Results}
\begin{figure*}
	\centering
        \includegraphics[width=0.8\textwidth]{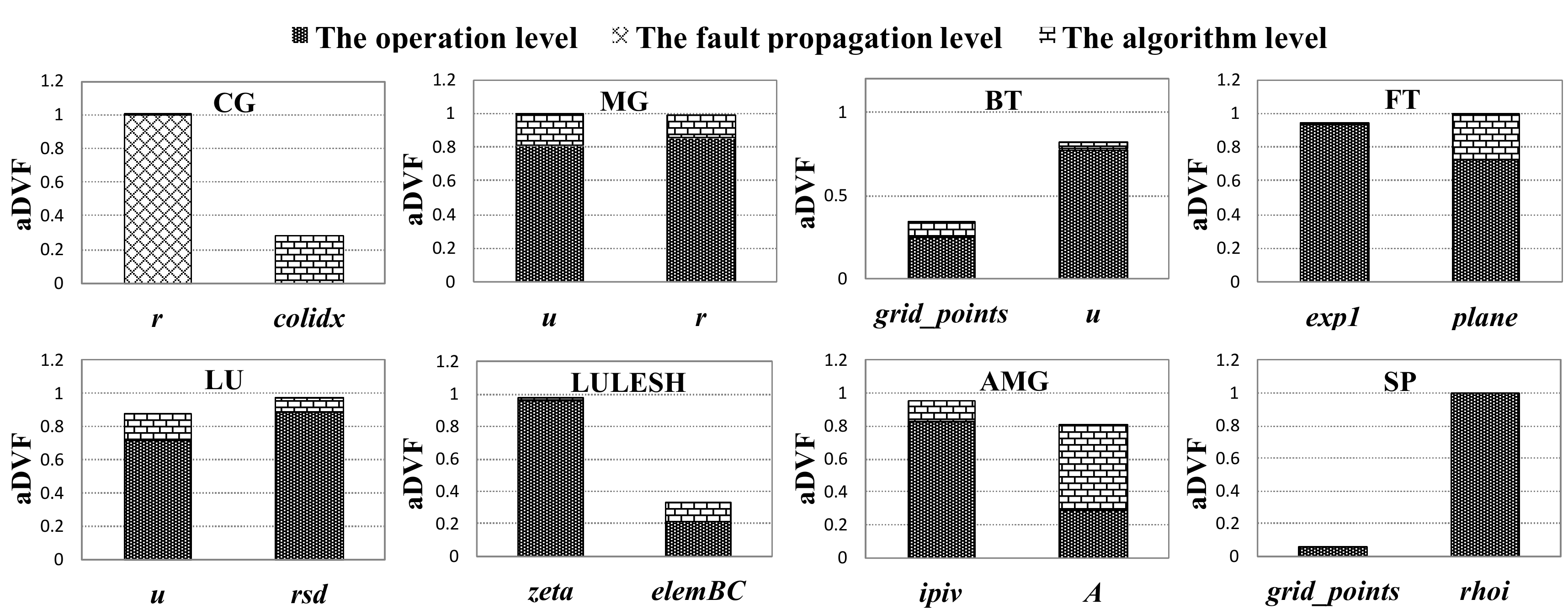}
%
        \vspace{-5pt}
        \caption{The breakdown of aDVF results based on the three level analysis. The $x$ axis is the data object name.}
        \vspace{-8pt}
        \label{fig:aDVF_3tiers_profiling}
\end{figure*}

\begin{figure*}
	\centering
	\includegraphics[width=0.8\textwidth]{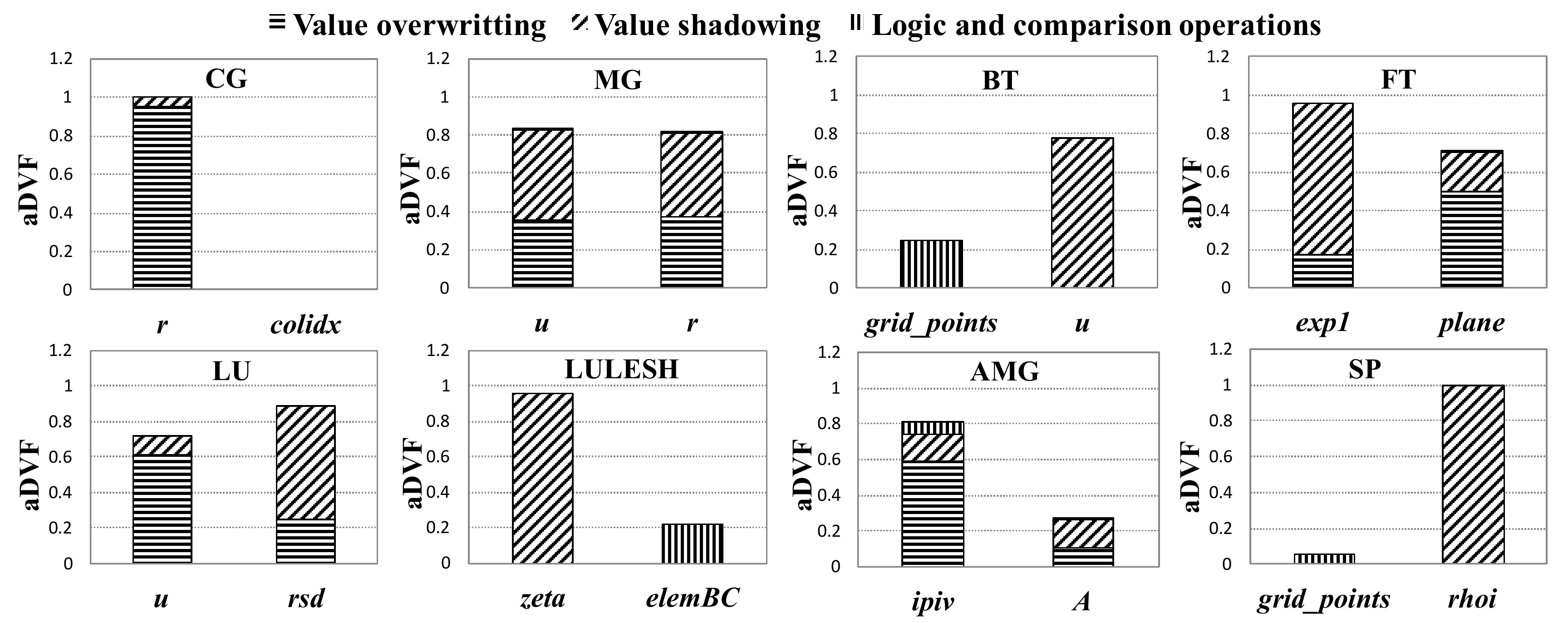}
	\vspace{-5pt}
	\caption{The breakdown of aDVF results based on the three classes of fault masking. The $x$ axis is the data object name. \textit{zeta} and \textit{elemBC} in LULESH are \textit{m\_delv\_zeta} and \textit{m\_elemBC} respectively.} 
	\vspace{-5pt}
	\label{fig:aDVF_3classes_profiling}
\end{figure*}

Figure~\ref{fig:aDVF_3tiers_profiling}
shows the aDVF results and breaks them down into the three levels 
(i.e., the operation-level, fault propagation level, and algorithm-level).
Figure~\ref{fig:aDVF_3classes_profiling} shows the 
results for the analyses at the levels of the operation and fault propagation,
and further breaks down the results into 
the three classes (i.e., the value overwriting, logical and comparison operations,
and value shadowing). 
We have multiple interesting findings from the results.

(1) Fault masking is common across benchmarks and applications.
Several data objects (e.g., $r$ in CG, and $exp1$ and $plane$ in FT)
have aDVF values close to 1 in Figure~\ref{fig:aDVF_3tiers_profiling}, 
which indicates that most of operations working on these data objects
have fault masking.
However, a couple of data objects have much less intensive fault masking.
For example, the aDVF value of $colidx$ in CG is 0.28 (Figure~\ref{fig:aDVF_3tiers_profiling}). 
Further study reveals that $colidx$ is an array to store column indexes of sparse matrices, and there is few operation-level or fault propagation-level fault masking  (Figure~\ref{fig:aDVF_3classes_profiling}).
The corruption of it can easily cause segmentation fault caught by the
algorithm-level analysis. 
$grid\_points$ in SP and BT also have a relatively small aDVF value (0.14 and 0.38 for SP and BT respectively in Figure~\ref{fig:aDVF_3tiers_profiling}).
Further study reveals that $grid\_points$ defines input problems for SP and BT. 
A small corruption of $grid\_points$ 
can easily cause major changes in computation
caught by the fault propagation analysis. 

The data object $u$ in BT also has a relatively small aDVF value (0.82 in Figure~\ref{fig:aDVF_3tiers_profiling}).
Further study reveals that $u$ is read-only in our target code region
for matrix factorization and Jacobian, neither of which is friendly
for fault masking.
Furthermore, the major fault masking for $u$ comes from value shadowing,
and value shadowing only happens in a couple of the least significant bits 
of the operands that reference $u$, which further reduces the value of aDVF.

(2) The data type is strongly correlated with fault masking.
Figure~\ref{fig:aDVF_3tiers_profiling} reveals that the integer data objects ($colidx$ in CG, $grid\_points$ in BT and SP, $m\_elemBC$ in LULESH) appear to be 
more sensitive to faults than the floating point data objects 
($u$ and $r$ in MG, $exp1$ and $plane$ in FT, $u$ and $rsd$ in LU, $m\_delv\_zeta$ in LULESH, and $rhoi$ in SP).
In HPC applications, the integer data objects are commonly employed to
define input problems and bound computation boundaries (e.g., $colidx$ in CG and $grid\_points$ in BT), 
or track computation status (e.g., $m\_elemBC$ in LULESH). Their corruption 
is very detrimental to the application correctness. 

(3) Operation-level fault masking is very common.
For many data objects, the operation-level fault masking contributes 
more than 70\% of the aDVF values. For $r$ in CG, $exp1$ in FT, and $rhoi$ in SP,
the contribution of the operation-level fault masking is close to 99\% (Figure~\ref{fig:aDVF_3tiers_profiling}).

Furthermore, the value shadowing is a very common operation level fault masking,
especially for floating point data objects (e.g., $u$ and $r$ in BT, $m\_delv\_zeta$ in LULESH, and $rhoi$ in SP in Figure~\ref{fig:aDVF_3classes_profiling}).
This finding has a very important indication for studying the application resilience.
In particular, the values of a data object can be different across different input problems. If the values of the data object are different, 
then the number of fault masking events due to the value shadowing will be different. 
Hence, we deduce that the application resilience
can be correlated with the input problems,
because of the correlation between the value shadowing and input problems. 
We must consider the input problems when studying the application resilience.
This conclusion is consistent with a very recent work~\cite{sc16:guo}.

(4) The contribution of the algorithm-level fault masking to the application resilience can be nontrivial.
For example, the algorithm-level fault masking contributes 19\% of the aDVF value for $u$ in MG and 27\% for $plane$ in FT (Figure~\ref{fig:aDVF_3tiers_profiling}).
The large contribution of algorithm-level fault masking in MG is consistent with
the results of existing work~\cite{mg_ics12}. 
For FT (particularly 3D FFT), the large contribution of algorithm-level fault masking in $plane$ (Figure~\ref{fig:aDVF_3tiers_profiling})
comes from frequent transpose and 1D FFT computations that average out 
or overwrite the data corruption.
CG, as an iterative solver, is known to have the algorithm-level fault masking
because of the iterative nature~\cite{2-shantharam2011characterizing}.
Interestingly, the algorithm-level fault masking in CG contributes most to the resilience of $colidx$ which is a vulnerable integer data object (Figure~\ref{fig:aDVF_3tiers_profiling}).


(5) Fault masking at the fault propagation level is small.
For all data objects, the contribution of the fault masking at the level of fault propagation is less than 5\% (Figure~\ref{fig:aDVF_3tiers_profiling}).
For 6 data objects ($r$ and $colidx$ in CG, $grid\_points$ and $u$ in BT, and 
$grid\_points$ and $rhoi$ in SP),  there is no fault masking at the level of fault propagation.
In combination with the finding 4, we conclude that once the fault
is propagated, it is difficult to mask it because of the contamination of
more data objects after fault propagation, and only the algorithm semantics can tolerate  propagated faults well. 

\begin{figure*}
	\centering
	\includegraphics[width=0.77\textheight, height=0.23\textheight]{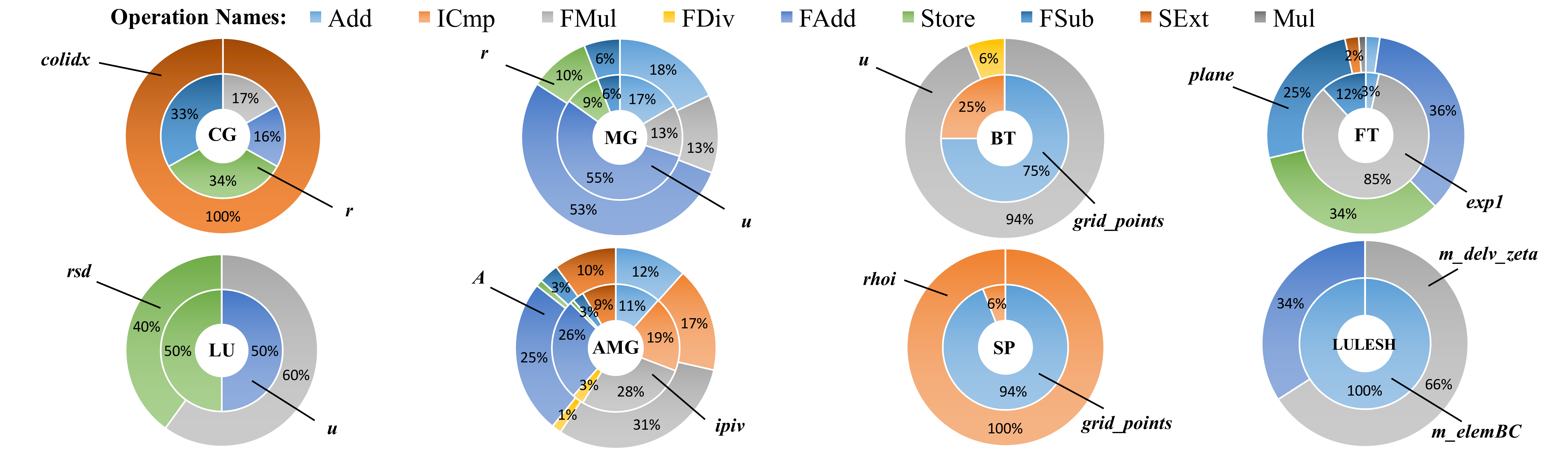}
	\vspace{-10pt}
	\caption{Breakdown of the aDVF results based on the analyses at the levels of operation and fault propagation}
    \vspace{-10pt}
	\label{fig:aDVF_fine_profiling}
\end{figure*}

(6) Fault masking by logical and comparison operations is small,
comparing with the contributions of value shadowing and overwriting (Figure~\ref{fig:aDVF_3classes_profiling}). 
Among all data objects, 
the logical and comparison operations in $grid\_points$ in BT contribute the most (25\% contribution in Figure~\ref{fig:aDVF_fine_profiling}), 
because of intensive ICmp operations (integer comparison). 

(7) The resilience varies across data objects. 
This fact is especially pronounced in two data objects $colidx$ and $r$ in CG (Figure~\ref{fig:aDVF_3tiers_profiling}).
 $colidx$ has aDVF much smaller than $r$, which means $colidx$ is much less resilient than $r$ (see finding 1 for a detailed analysis on $colidx$). 
Furthermore, $colidx$ and $r$ have different algorithm-level
fault masking (see finding 4 for a detailed analysis).

To further investigate the reasons for fault masking, 
we break down the aDVF results at the granularity of LLVM instructions,
based on the analyses at the levels of operation and fault propagation.
The results are shown in Figure~\ref{fig:aDVF_fine_profiling}.

(8) Arithmetic operations make a lot of contributions to fault masking.
For $r$ in CG, $u$ in BT, $plane$ and $exp1$ in FT, $m\_elemBC$ in LULESH, 
arithmetic operations (addition, multiplication, and division) contribute to almost 100\% of the fault masking (Figure~\ref{fig:aDVF_fine_profiling}).  

\subsection{Sensitivity Study}
\label{sec:eval_sen}
\begin{figure}
		\begin{center}
		\includegraphics[width=0.48\textwidth,height=0.11\textheight]{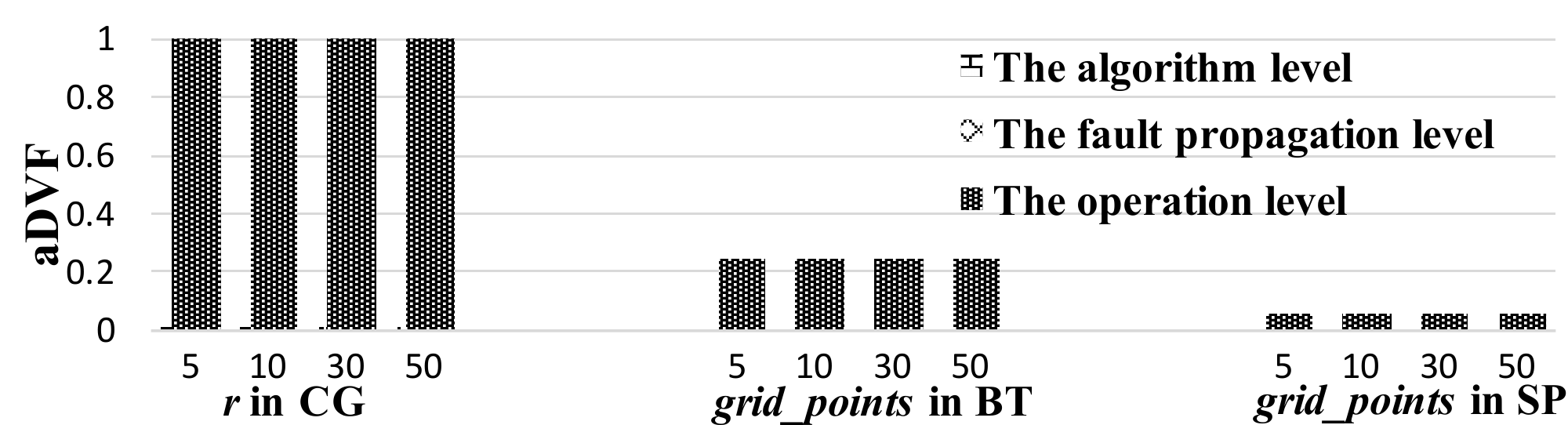}
		\vspace{-15pt}
		\caption{Sensitivity study for fault propagation threshold}
		\label{fig:sensitivity_error_propagation}
		\end{center}
\vspace{-15pt}
\end{figure}

ARAT uses 10 as the default fault propagation analysis threshold. 
The fault propagation analysis will not go beyond 10 operations. Instead,
we will use deterministic fault injection after 10 operations. 
In this section, we study the impact of this threshold on the modeling accuracy. We use a range of threshold values and examine how the aDVF value varies and whether
the identification of fault masking varies. 
Figure~\ref{fig:sensitivity_error_propagation} shows the results for 
multiple data objects in CG, BT, and SP.
We perform the sensitivity study for all 16 data objects.
Due to the page space limitation, we only show the results for three data objects,
but we summarize the sensitivity study results for all data objects in this section.

Our results reveal that the identification of fault masking by tracking fault propagation is not significantly 
affected by the fault propagation analysis threshold. Even if we use a rather large threshold (50), 
the variation of aDVF values is 4.48\% on average among all data objects,
and the variation at each of the three levels of analysis (the operation level, fault propagation level,  and algorithm level) is less than 5.2\% on average. 
In fact, using a threshold value of 5 is sufficiently accurate in most of the cases (14 out of 16 data objects).
This result is consistent with our finding 5 (i.e., fault masking at the fault propagation level is small). 
However, we do find a data object ($m\_elementBC$ in LULESH) 
showing relatively high-sensitive (up to 15\% variation) to the threshold. For this uncommon data object, using 50 as the fault propagation path is sufficient. 


\section{Case Study: Optimizing Fault Tolerance for Applications}
\label{sec:case_study}
aDVF and its analysis are widely applicable to a number of use cases, such as
code optimization and algorithm choice.
In this section, we study a case of using aDVF to help system designers to decide whether a specific application-level fault tolerance mechanism is helpful to improve the application resilience.

Application-level fault tolerance mechanisms, such as algorithm-based fault tolerance~\cite{abft_ecc:SC13, 21-chen2011algorithm, jcs13:wu, ics11:davies, hpdc13:davies, tc84_abft} and compiler-directed redundant execution~\cite{cgo05:reis, date05:hu, pact10:zhang, cgo07:wang, tr02:oh}, are extensively studied as a means to increase application resilience to faults. However, those application-level fault tolerance mechanisms can come with big performance and energy overheads 
(e.g., 35\% performance loss for dense matrix factorization in small scale deployments~\cite{ftfactor_ppopp12} and 41\% performance loss for compiler-directed instruction duplication~\cite{tr02:oh}). 
To justify the necessity of using these mechanisms, we must quantify how effectively those mechanisms improve the application resilience.
However, it is challenging to do so without a resilience metric and quantitative analysis method. 
With the introduction of aDVF, we can evaluate if the application resilience is effectively improved with fault tolerance mechanisms in place.

In this section, we focus on a specific application-level fault tolerance mechanism,
the algorithm-based fault tolerance (ABFT) for general matrix multiplication ($C=A \times B$)~\cite{jcs13:wu}.
This ABFT mechanism encodes matrices $A$, $B$ into a new form with checksums shown in the following equation, and protect $C$.
$e$ in the equation is an all-one column checksum vector. 

\scriptsize
\[ A^r:=\begin{bmatrix}A\\e^TA\end{bmatrix},
B^c:=\begin{bmatrix}B&Be\end{bmatrix} \]
\normalsize
To protect the result matrix $C$ from faults, instead of multiplying matrices $A$ by $B$, we use their checksum version.

\scriptsize
\[ A^rB^c=\begin{bmatrix} AB & ABe \\ e^TAB & e^TABe \end{bmatrix} =
\begin{bmatrix} C & Ce \\ e^TC & e^TCe \end{bmatrix} =: C^f \]
\normalsize
The new result matrix $C^f$ has extra checksum information, shown as above. The extra checksum information can be used to detect, locate, and recover fault during computation. 
If a fault occurs in an element of $C$, exactly one row and one column of the result
will not satisfy the checksum matrix definition. 
Then, leveraging either the row or column checksum, we are able to correct the faulty element. 

\begin{figure}
	\centering
	\includegraphics[width=0.49\textwidth, height=0.155\textheight]{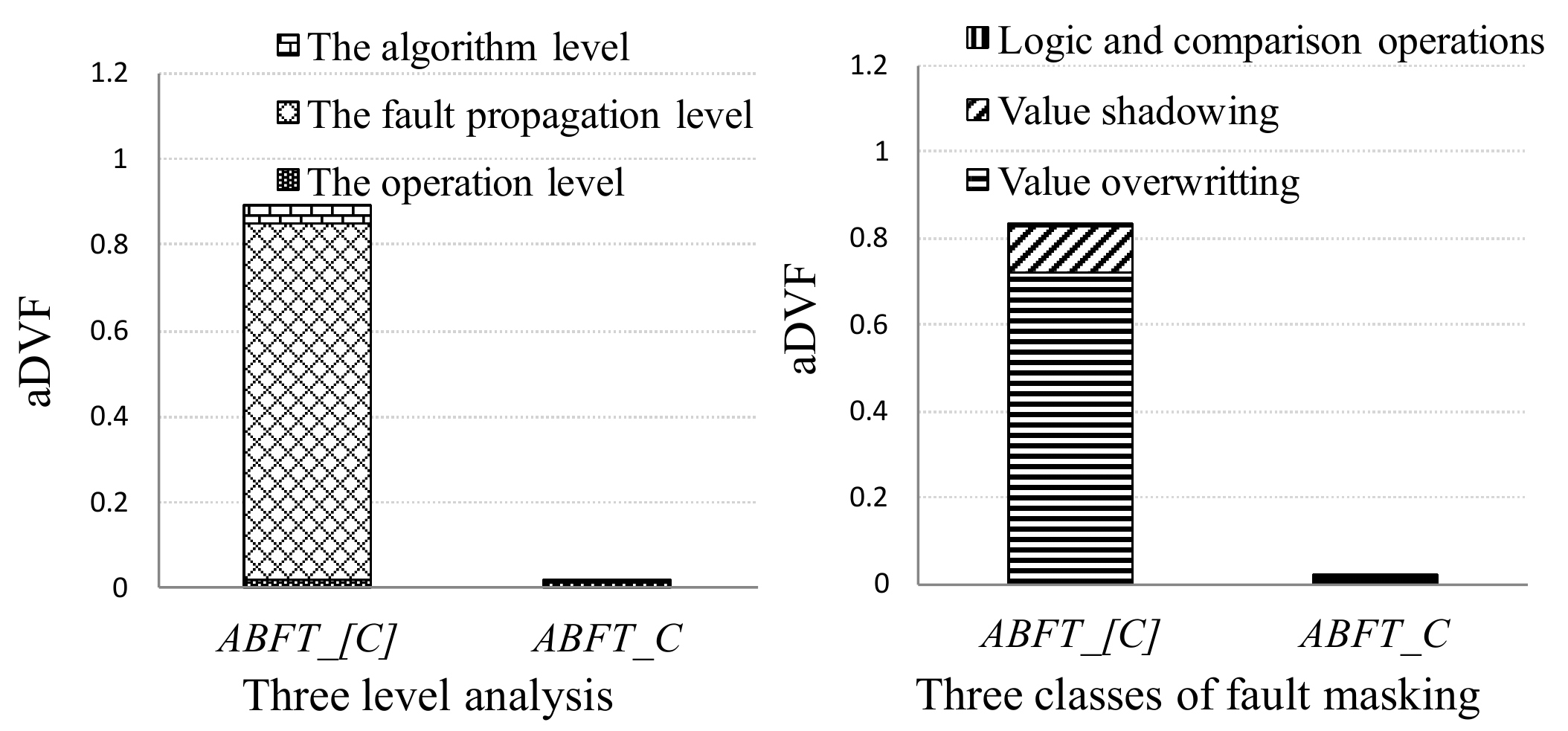}
	\caption{Using aDVF analysis to study the effectiveness of an ABFT for matrix multiplication (MM). ABFT\_C is MM without the protection of ABFT on $C$; ABFT\_[C] is MM with ABFT taking effect.}
	\label{fig:abft_advf}
    \vspace{-10pt}
\end{figure}

We apply the aDVF analysis on this ABFT, and the matrix $C$ is the target data object. We compare the aDVF values of $C$
with and without ABFT. Figure~\ref{fig:abft_advf} shows the results.
The figure shows that ABFT effectively improves the resilience of the matrix $C$: the aDVF value increases
from 0.0172 to 0.82. The improvement mostly comes from the value overwriting
at the fault propagation level.
This result is expected, because an element of $C$, once a fault happens in it, is not corrected by ABFT right way.
Instead, it will be corrected in a specific verification phase of ABFT.

Given the effectiveness of this ABFT, we further explore whether this ABFT can help us improve the resilience of data objects in an application, AMG.
AMG frequently employs matrix-vector multiplication. Given the fact that the vector can be treated as a special matrix, we can apply ABFT to protect 
the result vectors for those matrix-vector multiplications.
In particular, we protect a specific data object, the vector $r$, because this vector works as a result vector for 75\% of matrix-vector multiplications in AMG.
Using the vector $r$ as our target data object, we perform the aDVF analysis with and without ABFT.
We want to answer a question: Will using ABFT be an effective fault tolerance mechanism for AMG?

\begin{figure}
	\centering
	\includegraphics[width=0.49\textwidth,height=0.155\textheight]{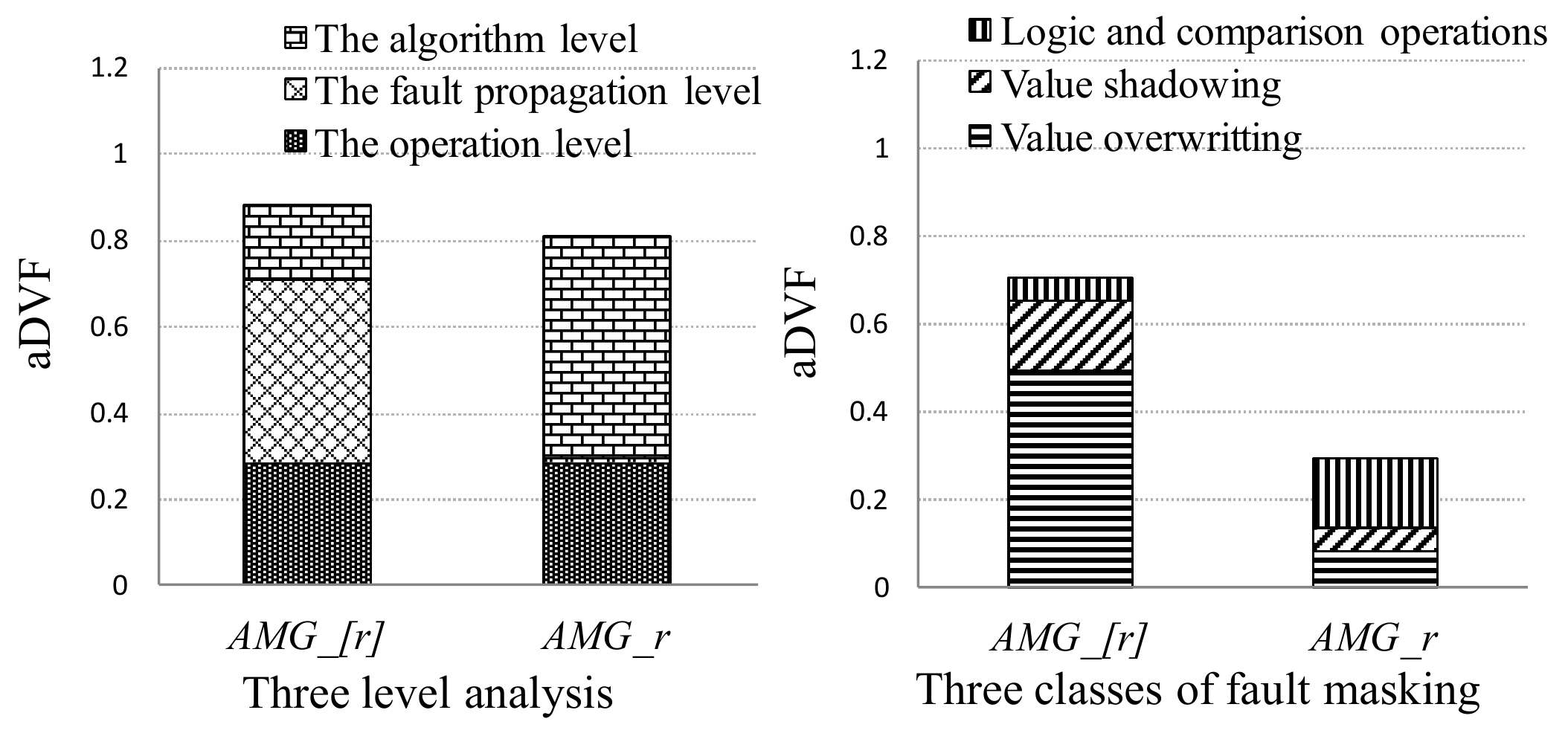}
	\caption{Using aDVF analysis to study the effectiveness of the ABFT for a data object $r$ in AMG. AMG\_r has no protection of ABFT; ABFT\_[r] has ABFT taking effect on $r$.}
	\label{fig:abft_advf_qbox}
    \vspace{-12pt}
\end{figure}

Figure~\ref{fig:abft_advf_qbox} shows the results. The figure reveals that using ABFT is not very helpful to improve
the resilience of the data object $r$ in AMG:
there is only a slightly change to the aDVF value.
After examining the AMG code, we found that the vector $r$ is involved in a generalized minimal residual method (GMRES).
This method approximates the solution
of a linear equation by a vector with minimal residual. 
The approximation nature of GMRES determines that the GMRES method itself can tolerate faults. 
For the vector $r$, most of the faults correctable by ABFT are also tolerable by GMRES.
Hence, ABFT is not very helpful to improve the application resilience.
Our aDVF analysis result is consistent with the above code analysis result.
Furthermore, Figure~\ref{fig:abft_advf_qbox} reveals that how faults are masked in AMG with and without ABFT. With ABFT,
most faults are masked at the fault propagation level, while without ABFT most faults are masked at the algorithm level. 

This case study is a clear demonstration of how powerful the aDVF analysis can help optimize
fault tolerance. Avoiding redundant protection as above will greatly improve performance and energy efficiency of HPC systems.
\vspace{-5pt}

\section{Related Work}
\begin{spacing}{0.9}
\textbf{Application-level random fault injection.}
Casa et al.~\cite{mg_ics12} 
study the resilience of an algebraic multi-grid solver 
by injecting faults into instructions' output based on LLVM. 
Similar work can be found in~\cite{europar14:calhoun, prdc13:sharma}.
Cher et al.~\cite{sc14:cher} employ a GDB-like debugging tool to corrupt register states.
Li et al.~\cite{bifit:sc12} build a binary instrumentation-based fault injection tool 
for random fault injection.
Shantharam et al.~\cite{2-shantharam2011characterizing} manually change the values of data objects 
to study the resilience of iterative methods.
Ashraf et al.~\cite{sc15:ashraf} and Wei et al.~\cite{dsn14:wei} also use LLVM-based tools to inject faults, but
they further introduce the functionality of tracking fault propagation. 
Xu et al.~\cite{dsn12:xu} and Hari et al.~\cite{asplos12:hari} aggressively employ static and dynamic program analyses to reduce the number of fault injection tests.
Their work 
still has randomness for fault injection.

\textbf{Resilience metrics.}
Architectural vulnerability factor (AVF) is a hardware-oriented metric
to quantify the probability of a fault in a hardware component resulting in
the incorrect final application outcome. It was first introduced  in~\cite{isca05:mukherjee, micro03:mukherjee},
and then attracted a series of follow-up work. This includes
statistical-based modeling techniques to accelerate AVF estimate~\cite{micro07:cho, hpca09:duan, mascots06:fu}, 
online AVF estimation~\cite{isca08:Li, isca07:soundararajan},  
and AVF analysis for spatial multi-bit faults~\cite{micro14:wilkening}.
Another metric, the program vulnerability factor (PVF) is based on AVF~\cite{hpca09:sridharan}, 
but eliminates microarchitecture effects. 

AVF calculation and its variants are highly hardware-oriented. 
In fact, AVF presents an aggregation effect of hardware and application, but is typically employed to evaluate the hardware vulnerability. AVF calculation usually requires detailed hardware simulations, and requires a
large number of simulations to derive insight into the impact of (micro)architectural events on AVF, which can be time-consuming.
Although the recent work based on statistical approaches improves evaluation speed~\cite{micro07:cho, hpca09:duan, mascots06:fu}, 
it limits modeling accuracy. 
AVF calculation does not consider application semantics, and hence can overestimate vulnerability.
Yu et al.~\cite{dvf_sc14} introduce a metric, DVF. DVF 
captures the effects of both application and hardware on the resilience of data
structures. 
In contrast to AVF and DVF, our metric, aDVF, is a highly 
application-oriented metric.
\end{spacing}

\vspace{-5pt}  
\section{Conclusions}
\label{sec:conclusions}
\begin{spacing}{0.9}
This paper introduces a new methodology to quantify the application resilience.
Different from the traditional random fault injection, 
our methodology employs a direct measurement of fault masking events inherent in applications. 
Based on our methodology, we introduce a new metric, a series of techniques, and a tool to analyze and identify error masking.
We hope that our methodology and tool can make the quantification of application resilience a common practice for evaluating applications in the future. 
\end{spacing}

\bibliographystyle{abbrv}
\bibliography{li,li_sc16_resilience_modeling}  

\begin{thebibliography}{10}
\bibitem{resilience_modeling:tr}
{(We omit paper details here due to double-blinded reviews) Application-Level
  Resilience Modeling}.
\end{thebibliography}

\end{document}